\documentclass[12pt,a4paper]{article}

\title{On nonlinear waves of the blood flow trough arteries}
\author{Elena Nikolova , Ivan P. Jordanov, Nikolay K. Vitanov}
\date{Institute of :echanics, Bulgarian Academy of Sciences, Akad. G. Bonchev Str., Bl. 4, 1113 Sofia, Bulgaria}
\begin{document}

\maketitle

\begin{abstract}
We discuss propagation of traveling waves in a blood filled elastic artery with an axially symmetric dilatation (an idealized aneurysm).
The processes in the injured artery are modelled by equations for the motion of the
wall of the artery and by equation for the motion of the fluid (the blood).
For the case when long-wave approximation holds the model equations are reduced
to a version of the perturbed Korteweg-deVries equation. Exact travelling-wave
solutions of this equation are obtained by the modified method of simplest
equation where the differential equation of Abel is used as a simplest equation.
A particular case of the obtained exact solution is discussed from the point of view of arterial mechanics.
\end{abstract}

\section{Introduction}
The theoretical investigation of pulse wave propagation in human arteries has a long history starting from ancient times until today. With the development of
methods for studying  nonlinear systems \cite{ss}--\cite{dv12}, great progress has been observed in understanding nonlinear waves in fluid-solid structures, such as blood-filled human arteries. The blood flow in arteries is complicated and because of this one often uses  simplifications in modeling of wall mechanics and blood hemodynamics. For example, it is known that arterial wall is an incompressible, anisotropic hyperelastic (nonlinear elastic) or viscoelastic material. But, because of the significant complexity of fluid-solid structure models under consideration of the wall anisotropy, in the most works the vessel wall is assumed to be isotropic. Moreover, taking into account also the wall thickness-vessel radius (diameter) ratio, the arteries are treated as straight incompressible, thick viscoelastic \cite{Demiray1998},\cite{Demiray1999} or elastic tubes \cite{DemirayDost1998}, as well as thin elastic \cite{Demiray2002}--\cite{Abulwafa2016} or viscoelastic  tubes \cite{Demiray2005},\cite{Shan2010}. Usually the arteries are modelled by circularly cylindrical long tubes with a constant cross-section.
On the other hand, although the blood is known to be an incompressible non-Newtonian fluid, in many studies (including the papers cited above), it was modeled as an incompressible inviscid fluid. In the above-mentioned papers, propagation of weakly nonlinear waves in the long-wave approximation is studied by employing different perturbation techniques. In dependence on the balance between nonlinearity and dispersion or dissipation, Korteweg-de Vries, Burgers and Korteweg-de Vries-Burgers type equations are derived as model equations. Shock or solitary type waves are obtained as solutions of the model equations  dependending on the  kind of model equation.
\par
 Over the past decade, however, the scientific efforts  have been concentrated on theoretical investigations of nonlinear wave propagation in an artery with a variable radius through the blood. Clearing how local imperfections appeared in an artery can disturb the blood flow can help in  predicting  the nature and main features of  various cardiovascular diseases, such as stenoses and aneurysms. In order to examine propagation of nonlinear waves in a stenosed artery, Tay and co-authors treated the artery as a homogeneous, isotropic and thin-walled elastic tube with an axially symmetric stenosis. The blood was modeled as  an incompressible inviscid fluid \cite{Tay2006}, Newtonian fluid with constant viscosity \cite{Tay2007},  and Newtonian fluid with variable viscosity \cite{Tay2008}. Using a specific perturbation method, in a long-wave approximation the authors obtained the forced KdV equation with variable coefficients \cite{Tay2006}, forced perturbed KdV equation with variable coefficients \cite{Tay2007}, and forced KdVB equation with variable coefficients as evolution equations \cite{Tay2008}. The same theoretical frame was used in \cite{Demiray2008}, \cite{Dimitrova2015} to examine nonlinear wave propagation in an artery with a local dilataion (an idealized aneurysm). Considering the dilated artery as a long inhomogeneous prestretched thin elastic tube with a local imperfection (presented at large by a unspecified function $f(z)$), and the blood as an incompressible inviscid fluid the authors reached again the forced KdV equation with variable coefficients. Apart from solitary propagation waves in such a system, in \cite{Dimitrova2015}, possibility of periodic waves was discussed at appropriate initial conditions. Motivated by investigations in \cite{Tay2006}--\cite{Dimitrova2015}, the main goal of this paper is to investigate effects of a particular arterial geometry and the blood characteristics on the propagation of nonlinear waves through an injured artery. For that purpose, we use a reductive perturbation method to obtain the nonlinear evolution equation. Exact solution of this equation is obtained by using the modified method of simplest equation. Recently, this method has been widely used to obtain general and particular solutions of economic, biological and physical models, represented by partial differential equations.
\par 
The paper is organized as follows. A brief description about the derivation of equations governing the blood flow trough a dilated artery will be presented in Sec. 2 and Sec. 3. In Sec. 4, a travelling wave solution of the evolution equation will be obtained. Numerical simulations of the solutions and main conclusions based on the obtained results will be summarized in Sec. 5 of the paper.

\section{Formulation of the equations for the tube and the fluid}

In order to derive the model of a blood-filled artery with an aneurysm we have to consider two types equations which represent (i) the motion of the arterial wall and (ii) the motion of the blood. To model such a medium we shall treat the artery as a thin-walled incompressible prestretched hyperelastic tube with a localized axially symmetric dilatation. We shall assume the blood to be an incompressible viscous fluid. A brief formulation of the above--mentioned equations follows in the next two subsections.

\subsection{Equation of the wall}
It is well-known, that for a healthy human, the systolic pressure is about 120 mm Hg and the diastolic pressure is 80 mm Hg. Thus, the arteries are initially subjected to a mean pressure, which is about 100 mm Hg. Moreover, the elastic arteries are initially  prestretched in an axial direction. This feature minimizes its axial deformations during the pressure cycle. For example, experimental studies declare that the longitudinal motion of arteries is very small \cite{Patel}, and it is due mainly to strong vascular tethering and partly to the predominantly circumferential orientation of the elastin and collagen fibers. Taking into account these observations, and following the methodology applied in \cite{Tay2006}--\cite{Demiray2008}, we consider the artery as a circularly cylindrical tube with radius $R_{0}$, assuming that such a tube is subjected to an initial axial stretch $\lambda_{z}$ and a uniform inner pressure $P_{0}^*(Z)$. Under the action of such a variable pressure the position vector of a generic point on the tube can be described by
\begin{equation}\label{eq1}
\underline{r_{0}}=[r_{0}+f^*(z^*)]e_{r}+z^*e_{z},\   z^*=\lambda_{z}Z^*
\end{equation}\\
where $e_{r}$ and $e_{z}$ are the unit basic vectors in the cylindrical polar coordinates, $r_{0}$ is the deformed radius at the origin of the
coordinate system, $Z^*$ is axial coordinate before the deformation, $z^*$ is the axial coordinate after static deformation and $f^*(z^*)$
is a function describing the dilatation geometry. In the literature, several analytical forms of idealized (axially symmetric) aneurysms exist \cite{Elger1996},\cite{Gopalakrishnan2014}. However, we shall specify the concrete form of $f^*(z^*)$ later. Upon the initial static deformation, we shall superimpose only a dynamical radial displacement $u^*(z^*,t^*)$, neglecting the contribution of axial displacement because of the experimental observations, given above. Then, the position vector
$r$ of a generic point on the tube can be expressed by
\begin{equation}\label{eq2}
\underline{r}=[r_{0}+f^*(z^*)+u^*]e_{r}+z^*e_{z}
\end{equation}
The arc-lengths along meridional and circumferential curves respectively, are:
\begin{equation}\label{eq3}
ds_{z}=[1+(f^{*\prime}+\frac{\partial{u^*}}{\partial{z^*}})^2]^{1/2}dz^*,\  ds_{\theta}=[r_{0}+f^*+u^*]d{\theta}
\end{equation}
In this way, the stretch ratios in the longitudinal and circumferential directions in final configuration are presented by
\begin{equation}\label{eq4}
\lambda_{1}=\lambda_{z}\Lambda, \ \lambda_{2}=\frac{1}{R_{0}}(r_{0}+f^*+u^*)
\end{equation}\\
where
\begin{equation}\label{eq5}
\Lambda=[1+(f^{*\prime}+\frac{\partial{u^*}}{\partial{z^*}})^2]^{1/2}
\end{equation}\\
The notation '$\prime$` denotes the differentiation of $f^*$ with respect to $z^*$. Then, the unit tangent vector $t$ along
the deformed meridional curve and the unit exterior normal vector $n$ to the deformed tube are expressed by
\begin{equation}\label{eq6}
t=\frac{(f^{*\prime}+\frac{\partial{u^*}}{\partial{z^*}})e_{r}+e_{z}}{\Lambda},\ \ n=\frac{e_{r}-(f^{*\prime}+\frac{\partial{u^*}}{\partial{z^*}})e_{z}}{\Lambda}
\end{equation}\\
According to the assumption made about material incompressibility the following restriction holds:
\begin{equation}\label{eq7}
h=\frac{H}{\lambda_{1}\lambda_{2}}
\end{equation}\\
where $H$ and $h$ are the wall thicknesses before and after deformation, respectively. The force, which acts on a small tube element placed between the planes $z^*=const$, $z^*+dz^*=const$, $\theta=const$ and $\theta+d\theta=const$, and thereby determines  its motion can be expressed in the following manner:
\begin{equation}\label{eq8}
F_{r}=-T_{2}\Lambda+\frac{\partial}{\partial{z^*}}\left\{\frac{(r_{0}+f^*+u^*)(f^{*\prime}+\partial{u^*}/\partial{z^*})T_{1}}{\Lambda}\right\}+P_{r}^*(r_{0}+f^*+u^*)\Lambda
\end{equation}\\
where $P_{r}^*$ is the fluid reaction force, which shall be specified later, and $T_{1}$ and $T_{2}$ are the tensions in longitudinal and circumferential directions, respectively. For hyperelastic materials, the corresponding tensions have the form:
\begin{equation}\label{eq9}
T_{1}=\frac{\mu H}{\lambda_{2}}\frac{\partial{\Pi}}{\partial{\lambda_{1}}},\ T_{2}=\frac{\mu H}{\lambda_{1}}\frac{\partial{\Pi}}{\partial{\lambda_{2}}}
\end{equation}\\
where $\Pi$ is the strain energy density function of wall material and $\mu^*$ is the material shear modulus. Although the elastic properties of an injured wall section differ from those of the healthy part, here, we assume that the wall is homogeneous, i.e. $\mu^*$ is a constant through the axis $z$. Finally, according to the second Newton's law, the equation of radial motion of a small tube element placed between the planes $z^*=const$, $z^*+dz^*=const$, $\theta=const$ and
$\theta+d\theta=const$ obtains the form:
\begin{equation}\label{eq10}
-\frac{\mu}{\lambda_{z}}\frac{\partial{\Pi}}{\partial{\lambda_{2}}}+\mu R_{0}\frac{\partial}{\partial{z^*}}\left\{\frac{(f^{*\prime}+\partial{u^*}/\partial{z^*})}{\Lambda}\frac{\partial{\Pi}}{\partial{\lambda_{1}}}\right\}+\frac{P_{r}^*}{H}(r_{0}+f^*+u^*)\Lambda=\rho_{0}\frac{R_{0}}{\lambda_{z}}\frac{\partial^2{u^*}}{\partial{t^{*2}}}
\end{equation}\\
where $\rho_{0}$ is the mass density of the tube material. 
\subsection{Equation of the fluid} 
Experimental studies over many years demonstrated that blood behaves as an incompressible non-Newtonian fluid because it consists of a suspension of cell formed elements in a liquid well-known as blood plasma. However, in the larger arteries (with a vessel radius  larger than 1 mm) it is plausible to assume that the blood has an approximately constant viscosity, because the vessel diameters are essentially larger than the individual cell diameters. Thus, in such vessels the non-Newtonian behavior becomes insignificant and the blood can be considered as a Newtonian fluid. Then, analogically to \cite{Tay2007} we present the axially symmetric motion of blood in polar coordinates as follows
\begin{equation}\label{eq11}
\frac{\partial{V_{r}^*}}{\partial{t^*}}+V_{r}^*\frac{\partial{V_{r}^*}}{\partial{r}}+V_{z}^*\frac{\partial{V_{r}^*}}{\partial{z^*}}+\frac{1}{\rho_{f}}\frac{\partial{\overline{P}}}{\partial{r}}-\frac{\mu_{f}}{\rho_{f}}(\frac{\partial^2{V_{r}^*}}{\partial{r^2}}+\frac{1}{r}\frac{\partial{V_{r}^*}}{\partial{r}}-\frac{V_{r}^*}{r^2}+\frac{\partial^2{V_{r}^*}}{\partial{z^{*2}}})=0
\end{equation}
\begin{equation}\label{eq12}
\frac{\partial{V_{z}^*}}{\partial{t^*}}+V_{r}^*\frac{\partial{V_{z}^*}}{\partial{r}}+V_{z}^*\frac{\partial{V_{z}^*}}{\partial{z^*}}+\frac{1}{\rho_{f}}\frac{\partial{\overline{P}}}{\partial{z^*}}-\frac{\mu_{f}}{\rho_{f}}(\frac{\partial^2{V_{z}^*}}{\partial{r^2}}+\frac{1}{r}\frac{\partial{V_{z}^*}}{\partial{r}}+\frac{\partial^2{V_{r}^*}}{\partial{z^{*2}}})=0
\end{equation}
\begin{equation}\label{eq13}
\frac{\partial{V_{r}^*}}{\partial{r}}+\frac{V_{r}^*}{r}+\frac{\partial{V_{z}^*}}{\partial{z^*}}=0
\end{equation}\\
where $V_{r}^*$ , $V_{z}^*$ are the radial and the axial velocity components, respectively, $\rho_{f}$ is the fluid mass density, $\overline{P}$ is the pressure function, and $\mu_{f}$ is the viscosity of the fluid. These field equations have to satisfy the following boundary conditions:
\begin{equation}\label{eq14}
V_{r}^*\mid_{r=r_{f}}=\frac{\partial{u^*}}{\partial{t^*}},\ \overline{P}\mid_{r=r_{f}}=P_{r}^*
\end{equation}
Next, we use so called 'hydraulic approximation`, and we apply an averaging procedure with respect to the cross-sectional area to Eqs (\ref{eq11})--(\ref{eq13}). 
Then, we have
\begin{equation}\label{eq15}
\frac{\partial{A^*}}{\partial{t^*}}+\frac{\partial}{\partial{z^*}}(A^* \omega^*)=0
\end{equation}
\begin{equation}\label{eq16}
\frac{\partial{\omega^*}}{\partial{t^*}}+\omega^*\frac{\partial{\omega^*}}{\partial{z^*}}+\frac{1}{\rho_{f}}\frac{\partial{P^*}}{\partial{z^*}}-\frac{\mu_{f}}{\rho_{f}}(\frac{\partial^2{\omega^*}}{\partial{z^{*2}}}-\frac{8 \omega^*}{r_{f}^2})=0
\end{equation}
where $A^*$ denotes the inner cross-sectional area, i.e., $A^* = \pi r_{f}^2$ as $r_{f}=r_{0}+f^*(z^*)+u^*$ is the final radius of the tube after deformation, $\omega^*$ is the averaged axial velocity and $P^*$ is the averaged pressure of the fluid. We obtain Eq. (\ref{eq16}) by the help of the relationships 
\begin{equation}\label{eq17}
A^* \omega^*=2\pi\int_{0}^{r_{f}}{rV_{z}^*dr},\ \ A^*P^*=2\pi\int_{0}^{r_{f}}{r\overline{P}dr}
\end{equation}
Similarly to \cite{Tay2007}, we use the following assumption \cite{Prandtl1957}
\begin{equation}\label{eq18}
A (\omega^*)^2=2 \pi \int_{0}^{r_{f}}{rV_{z}^{*2}dr},\ \ \frac{2\mu_{f}}{\rho_{f}r_{f}}\frac{\partial{V_{z}^*}}{\partial{r}}\mid_{r=r_{f}}=-\frac{8 \mu_{f} \omega^*}{\rho_{f} r_{f}^2}
\end{equation}\\
We note, that the above assumption is valid for blood flows through elastic arteries in the cases when the axial flow prevails over the radial one and the amplitude of diameter disturbance (variation) is small in comparison with its steady-state value. We introduce the expression  of $A^*$ into (15), and the following field equation is derived
\begin{equation}\label{eq19}
2\frac{\partial{u^*}}{\partial{t^*}}+2 \omega^*[f^{*\prime}+\frac{\partial{u^*}}{\partial{z^*}}]+[r_{0}+f^*(z^*)+u^*]\frac{\partial{\omega^*}}{\partial{z^*}}=0
\end{equation}\\
Eqs (\ref{eq16}) and (\ref{eq19}) give only two relations to determine the unknown variables $\omega^*$ and $u^*$ and $P^*$. The system can be closed by the following boundary condition: 
\begin{equation}\label{eq20}
P_{r}^*=\frac{1}{\Lambda}[P^*-\frac{4 \mu_{f}(f^{*\prime}+\partial{u^*}/\partial{z^*})\omega^*}{r_{0}+f^*(z^*)+u^*}]
\end{equation}\\
Next, we introduce the following non-dimensional quantities:
\begin{eqnarray}\label{eq21}
t^*=(\frac{R_{0}}{c_{0}})t,\ \ z^*=R_{0}z,\ \ u^*=R_{0}u,\ \ f^*=R_{0}f,\ \ \omega^*=c_{0}\omega,\ \ \mu_{f}=c_{0}R_{0}\rho_{f}\nu,\nonumber\\
P^*=\rho_{f}c_{0}^{2}p,\ \ r_{0}=R_{0}\lambda_{\theta},\ \ c_{0}^2=\frac{\alpha H}{\rho_{f} R_{0}},\ \ m=\frac{\rho_{0} H}{\rho_{f}R_{0}}
\end{eqnarray}\\
where $\lambda_{\theta} = r_{0}/R_{0}$ is the initial circumferential stretch ratio, and $c_{0}$ is the Moens-Korteweg wave speed.
Introducing (\ref{eq21}) into the equations (\ref{eq19}), (\ref{eq16}) and (\ref{eq20}), the following non-dimensional equations are obtained
\begin{equation}\label{eq22}
2\frac{\partial{u}}{\partial{t}}+2 \omega[f^{\prime}+\frac{\partial{u}}{\partial{z}}]+[\lambda_{\theta}+f(z)+u]\frac{\partial{\omega}}{\partial{z}}=0
\end{equation}
\begin{equation}\label{eq23}			
\frac{\partial{\omega}}{\partial{t}}+\omega\frac{\partial{\omega}}{\partial{z}}+\frac{\partial{p}}{\partial{z}}-\nu[\frac{\partial^2{\omega}}{\partial{z^{2}}}-\frac{8 \omega}{(\lambda_{\theta}+f(z)+u)^2}]=0
\end{equation}
\begin{eqnarray}\label{eq24}			
p=\frac{m}{\lambda_{z}(\lambda_{\theta}+f(z)+u)}\frac{\partial^2{u}}{\partial{t^2}}+\frac{1}{\lambda_{z}(\lambda_{\theta}+f(z)+u)}\frac{\partial{\Pi}}{\partial{\lambda_{2}}}-\nonumber \\
\frac{1}{(\lambda_{\theta}+f(z)+u)}\frac{\partial}{\partial{z}}(\frac{f^\prime+\partial{u}/\partial{z}}{\Lambda})\frac{\partial{\Pi}}{\partial{\lambda_{1}}}+\frac{4\nu(f^\prime+\partial(u)/\partial(z))\omega}{\lambda_{\theta}+f(z)+u}
\end{eqnarray}
Eqs (\ref{eq22})--(\ref{eq24}) give sufficient conditions to determine the variables $u, \omega$ and $p$.

\section{Derivation of the evolution equation in a long--wave approximation}
In this section we shall use the long--wave approximation to employ the reductive perturbation method \cite{Jeffrey1981} for studying the propagation of small-but-finite amplitude waves in a fluid-solid structure system, whose dimensionless governing equations are presented by (\ref{eq22})--(\ref{eq24}). In the long-wave limit, it is assumed that the variation of radius along the axial coordinate is small compared with the wave length. Because this condition is valid for large arteries, next we shall treat the problem as boundary value problem. We linearize Eqs (22)--(24), and search for their harmonic kind solution. As a result, the following dispersion relation is derived
\begin{equation}\label{eq25}
(2+\frac{m k^2}{\lambda_{z}})\upsilon^2-\lambda_{\theta}k^2(\gamma_{1}+\gamma_{0}k^2)=0
\end{equation} 
where $\upsilon$ is the angular frequency, $k$ is the wave number, $\gamma_{0}$  and $\gamma_{1}$ are some constants, which will be specified bellow. Expanding $k=k(\upsilon)$ into a power series around $\upsilon=0$, we obtain
\begin{equation}\label{eq26}
k=d_{1}\upsilon(1+d_{2}\upsilon^2)
\end{equation}
where
\begin{equation}\label{eq27}
d_{1}=(\frac{2}{\lambda_{\theta}\gamma_{1}})^{1/2},\ d_{2}=\frac{1}{\lambda_{\theta}\gamma_{1}}(\frac{m}{2\lambda_{z}}-\frac{\gamma_{0}}{\gamma_{1}})
\end{equation}
Thus, assuming $\epsilon=O(\upsilon^2)$  for the boundary value problems, it will be appropriate to introduce the following type of stretched
coordinates
\begin{equation}\label{eq28}				
\xi=\epsilon^{1/2}(z-ct),\ \ \tau=\epsilon^{3/2}z
\end{equation}
where $\epsilon$ is a small parameter measuring the weakness of nonlinearity and dispersion, and and $c$ is the scale parameter
which will be determined from the solution. Then, $z=\epsilon^{-3/2} \tau$, and the variables $u, \omega$ and $p$ are functions of the slow variables $(\xi,\tau)$ as well as the fast variables $(z, t)$. Taking into account the effect of dilatation, the function $f(z)$ is presented in the form $f(z)=\epsilon h(\tau)$. In addition, taking into account the effect of viscosity, the order of viscosity is assumed to be $O(3/2)$, i.e. $\nu=\epsilon^{3/2}\overline{\nu}$. The last assumption corresponds to the case of laminar flow, which is permissible  in modeling the blood flow in large arteries. According to the the long-time limit and reductive perturbation method, we expand the variables, presented in (\ref{eq22})--(\ref{eq24}) into asymptotic series as follows:
\begin{equation}\label{eq29}			
u=\epsilon u_{1}+\epsilon^2 u_{2}+....,\ \omega=\epsilon \omega_{1}+\epsilon^2 \omega_{2}+....,\ p=p_{0}+\epsilon p_{1}+\epsilon^2 p_{2}+...,
\end{equation}
Taking into account these expansions, the other quantities in (\ref{eq24}) are expanded also in terms of a small parameter $\epsilon$ in the following manner:
\begin{equation}\label{eq30}
\lambda_{1}\cong \lambda_{z},\ \lambda_{2}=\lambda_{\theta}+\epsilon (u_{1}+h(\tau))+\epsilon^2 (u_{2}+(u_{1}+h(\tau))^2)+...
\end{equation}
\begin{eqnarray}\label{eq31}
[\lambda_{\theta}+\epsilon h(\tau)+u]^{-1}=\frac{1}{\lambda_{\theta}}(1-\frac{(u_{1}+h(\tau))}{\lambda_{\theta}}\epsilon+(\frac{u_{2}}{\lambda_{\theta}}+\frac{(u_{1}+h(\tau))^2)}{\lambda_{\theta}^2})\epsilon^2-...),\nonumber \\
\frac{1}{\lambda_{\theta}\lambda_{z}}\frac{\partial{\Pi}}{\partial{\lambda_{z}}}=\gamma_{0},
\frac{1}{\lambda_{\theta}\lambda_{z}}\frac{\partial{W}}{\partial{\lambda_{\theta}}}=\beta_{0}+\beta_{1}(u_{1}+h)\epsilon+(\beta_{1}u_{2}+\beta_{2}(u_{1}+h)^2)\epsilon^2+..
\end{eqnarray}
where
\begin{equation}\label{eq32}	
\beta_{0}=\frac{1}{\lambda_{\theta}\lambda_{z}}\frac{\partial{\Pi}}{\partial{\lambda_{\theta}}},\ \ \beta_{1}=\frac{1}{\lambda_{\theta}\lambda_{z}}\frac{\partial^2{\Pi}}{\partial{\lambda_{\theta}^2}},\ \ \beta_{2}=\frac{1}{2 \lambda_{\theta}\lambda_{z}}\frac{\partial^3{\Pi}}{\partial{\lambda_{\theta}^3}},\ \ 
\end{equation}
Substituting (\ref{eq28})--(\ref{eq31}) into Eqs. (\ref{eq22})--(\ref{eq24}), we obtain the following differential sets:\\

\textit{O ($\epsilon$) equations}

\begin{equation}\label{eq33}				
-2c\frac{\partial{u_{1}}}{\partial{\xi}}+\lambda_{\theta}\frac{\partial{\omega_{1}}}{\partial{\xi}}=0,\\  -c\frac{\partial{\omega_{1}}}{\partial{\xi}}+\frac{\partial{p_{1}}}{\partial{\xi}}=0,\\\   p_{1}=\gamma_{1}(u_{1}+h(\tau))
\end{equation}

\textit{O ($\epsilon^2$) equations}\\

\begin{eqnarray}\label{eq34}
-2c \frac{\partial{u_{2}}}{\partial{\xi}}+2\omega_{1} \frac{\partial{u_{1}}}{\partial{\xi}}+\lambda_{\theta}\frac{\partial{\omega_{2}}}{\partial{\xi}}+[u_{1}+h]\frac{\partial{\omega_{1}}}{\partial{\xi}}+\lambda_{\theta}\frac{\partial{\omega_{1}}}{\partial{\tau}}=0\nonumber\\
-c\frac{\partial{\omega_{2}}}{\partial{\xi}}+\omega_{1}\frac{\partial{\omega_{1}}}{\partial{\xi}}+\frac{\partial{p_{2}}}{\partial{\xi}}+\frac{\partial{p_{1}}}{\partial{\tau}}+\frac{8\overline{\nu} \omega_{1}}{\lambda_{\theta}^2}=0\\
p_{2}=(\frac{m c^2}{\lambda_{\theta}\lambda{z}}-\gamma_{0})\frac{\partial^2{u_{1}}}{\partial{\xi}^2}+\gamma_{1} u_{2}+\gamma_{2}u_{1}^2+\gamma_{3}(\tau)u_{1}+\gamma_{4}(\tau)\nonumber
\end{eqnarray}
where
\begin{equation}\label{eq35}				
\ \gamma_{1}=\beta_{1}-\frac{\beta_{0}}{\lambda_{\theta}},\ \gamma_{2}=\beta_{2}-\frac{\beta_{1}}{\lambda_{\theta}},\ \gamma_{3}(\tau)=2 h(\tau)\gamma_{2},\ \gamma_{4}(\tau)=h^2(\tau)\gamma_{2}
\end{equation}
Solving Eqs (\ref{eq33}), we obtain
\begin{equation}\label{eq36}				
u_{1}=U(\xi, \tau),\ \ \omega_{1}=\frac{2c}{\lambda_{\theta}}U, \ \ p_{1}=\frac{2c^2}{\lambda_{\theta}}U+\gamma_{1} h(\tau)
\end{equation}
where $U(\xi, \tau)$ is an unknown function whose governing equation shall be obtained below and $c$ is the phase velocity in a long-wave approximation. Comparing the pressure terms $p_{1}$ expressed by (\ref{eq33}) and (\ref{eq36}), in order to have a non-zero solution for $U(\xi, \tau)$, the condition $\gamma_{1}=2 c^2/\lambda_{\theta}$ must be hold. We introduce (\ref{eq36}) in (\ref{eq34}), and obtain

\begin{equation}	\label{eq37}		
-2 c \frac{\partial{u_{2}}}{\partial{\xi}}+\frac{4c}{\lambda_{\theta}}U\frac{\partial{U}}{\partial{\xi}}+\lambda_{\theta}\frac{\partial{\omega_{2}}}{\partial{\xi}}+2c \frac{\partial{U}}{\partial{\tau}}+\frac{2c}{\lambda_{\theta}}(U+h)\frac{\partial{U}}{\partial{\xi}}=0
\end{equation}
\begin{equation}\label{eq38}			
-c \frac{\partial{\omega_{2}}}{\partial{\xi}}+\frac{4c^2}{\lambda_{\theta}^2}U\frac{\partial{U}}{\partial{\xi}}+\frac{2c^2}{\lambda_{\theta}}\frac{\partial{U}}{\partial{\tau}} +\gamma_{1} h^{\prime}+\frac{\partial{p_{2}}}{\partial{\xi}}+\frac{16 c \overline{\nu}}{\lambda_{\theta}^3}U=0
\end{equation}
\begin{equation}\label{eq39}	
p_{2}=(\frac{m c^2}{\lambda_{\theta}\lambda{z}}-\gamma_{0})\frac{\partial^2{U}}{\partial{\xi}^2}+\gamma_{1} u_{2}+\gamma_{2}U^2+\gamma_{3}(\tau)U+\gamma_{4}(\tau)
\end{equation}
By replacing (\ref{eq39}) into (\ref{eq38}), and eliminating $\omega_{2}$ between (\ref{eq37}) and (\ref{eq38}), we derive the evolution equation in a form:
\begin{equation}	\label{eq40}		
\frac{\partial{U}}{\partial{\tau}}+\mu_{1}U\frac{\partial{U}}{\partial{\xi}}+\mu_{2}\frac{\partial^3{U}}{\partial{\xi}^3}+\mu_{3}U+\mu_{4}(\tau)\frac{\partial{U}}{\partial{\xi}}+\mu(\tau)=0
\end{equation}
where
\begin{equation}\label{eq41}			
\mu_{1}=\frac{5}{2\lambda_{\theta}}+\frac{\gamma_{2}}{\gamma_{1}},\ \ \mu_{2}=\frac{m}{4 \lambda_{z}}-\frac{\gamma_{0}}{2 \gamma_{1}},\ \ \mu_{3}=\frac{4 \overline{\nu}}{c \lambda_{\theta}^2},\ \ \mu_{4}(\tau)=h(\tau)(\frac{1}{2 \lambda_{\theta}}+\frac{\gamma_{2}}{\gamma_{1}}),\ \ \mu(\tau)=\frac{1}{2} h^{\prime}(\tau)
\end{equation} 
Finally, we have to objectify the idealized aneurysm shape. For an idealized abdominal aortic aneurysm (AAA), $h(\tau)=\delta exp(\frac{-\tau^2}{2 L^2})$, where $\delta$ is the aneurysm height, i.e. $\delta=r_{max}-r_{0}$, and $L$ is the aneurysm length \cite{Gopalakrishnan2014}. In order to normalize these geometric quantities, we non-dimensionalize  $\delta$ by the inlet radius (diameter). Then, the non-dimensional coefficient can be presented by $\delta^{'}=DI-1$, where $DI=2r_{max}/2r_{0}=D_{max}/D_{0}$ is a geometric measure of AAA, which is known as a diameter index or a dilatation index \cite{Raut2013},\cite{Kleinstreuer2006}. In the same manner, the aneurysm length is normalized by the maximum aneurysm diameter ($D_{max}$), i.e. $SI=L/D_{max}$, where $SI$ is a ratio, which is known as a sacular index of AAA \cite{Raut2013}, \cite{Kleinstreuer2006}. For AAAs, $D_{max}$ varies from 3 cm to 8.5 cm, and $L$ varies from 5 cm to 10 cm.

\section{Analytical solution for the nonlinear evolution equation: Application of the modified method of simplest equation}
In this section we shall derive a progressive wave solution for the variable coefficients evolution equation, presented by (40). For this purpose, firstly we reduce it to an equation with constant coefficients, following two basic steps.\\
Step I: Let us introduce $U(\xi,\tau)=W(\xi,\tau)+\phi(\tau),$ where the function $\phi(\tau)$ is defined so that the new equation to be homogeneous, i.e. $\mu(\tau) = 0$. Then, we come from the linear equation $\phi^\prime (\tau)+\mu_3 \phi(\tau)=-\mu(\tau)$, which has a general solution:$$\phi(\tau)=\exp\big(-\int{\mu_3 d\tau}\big)\big[C(\xi)-\int{\mu(\tau)exp\big(\int{\mu_3 d\tau}\big)d\tau}\big].$$ At $C(\xi)=0$, i.e. $\phi(\tau)=\exp(-\mu_3 \tau)\big[-\int{\mu(\tau)exp(\mu_3 \tau)d\tau}\big]$ (at appropriate initial conditions) we make the substitution $$U(\xi,\tau)=W(\xi,\tau)-\exp(-\mu_3 \tau)\big[\int{\mu(\tau)\exp(\mu_3 \tau)d\tau}\big],$$ and obtain the evolution equation in the form:
\begin{equation}\label{eq42}
\frac{\partial{W}}{\partial{\tau}}+\mu_{1}W\frac{\partial{W}}{\partial{\xi}}+\mu_{2}\frac{\partial^3{W}}{\partial{\xi}^3}+\mu_{3}W+\mu_{4}(\tau)\frac{\partial{W}}{\partial{\xi}}+\mu_1 \phi(\tau)\frac{\partial{W}}{\partial{\xi}}=0.
\end{equation}
Step II: Let us change the variables using substitution:$$\tau=\tau(t^{\prime},x)=t^{\prime}, \xi=\xi(t^{\prime},x)=x+\psi(t^{\prime}).$$
We shall define the function $\psi$, so that the coefficient in front of $\frac{\partial{W}}{\partial{x}}$ to be zero. 
Then we obtain the equation
\begin{equation} \label{eq43}
\frac{d\psi}{dt^{\prime}}=\mu_1 \phi(t^{\prime})-\mu_{4} (t^{\prime}),
\end{equation} 
which has a general solution:
\begin{equation}\nonumber\\
\psi=\int{\big\{\big[exp(-\mu_3 \tau)\int{\mu(\tau)exp(\mu_3\tau)d\tau}\big]-\mu_4 (t)\big\}dt}+C_1 (x)
\end{equation}
 At $C_1 (x)=0$, the evolution equation is reduced to the  perturbed Korteweg-de
Vries equation:
\begin{equation}\label{eq44}
\frac{\partial{W}}{\partial{t^{\prime}}}+\mu_{1}W\frac{\partial{W}}{\partial{x}}+\mu_{2}\frac{\partial^3{W}}{\partial{x^3}}+\mu_{3}W=0
\end{equation}
In the last equation $\mu_{1}$ describes the nonlinearity, $\mu_{2}$ the dispersion and $\mu_{3}$ the dissipation. In principle, the  perturbed Korteweg-de
Vries equation is analytically unsolvable in the classical meaning. Thus, we shall consider the  unperturbed Korteweg-de
Vries equation, assuming that the dissipation is very small. When $\mu_{3}\approx{0}$, Eq. (44) is reduced to the generalized Korteweg-de Vries equation, that is 
\begin{equation}\label{eq45}
\frac{\partial{W}}{\partial{t^{\prime}}}+\mu_{1}W\frac{\partial{W}}{\partial{x}}+\mu_{2}\frac{\partial^3{W}}{\partial{x^3}}=0
\end{equation}
Although the exact analytical solution of this equation is well-known, here, we shall present another analytical form of the KdV equation applying the modified method of simplest equation to (\ref{eq45}). Modified method of simplest equation is a version of the method of simplest equation
\cite{k1,k2}. Modified method of simplest equation \cite{v1}-\cite{v4} was
used for obtaining exact analytical solution of numerous nonlinear PDes
\cite{v5} - \cite{v10}. The short description of the modified method of simplest equation is as follows. First of all by means 
of an appropriate ansatz (for an example the traveling-wave ansatz) the solved  of nonlinear partial differential equation for the unknown function $u$ is reduced to a nonlinear
ordinary differential equation that includes $\eta$ and its derivatives with respect to the
traveling wave coordinate $\zeta$
\begin{equation}\label{eq46}
P \left( \eta, \eta_{\zeta},\eta_{\zeta \zeta},\dots \right) = 0
\end{equation}
Then the finite-series solution 
\begin{equation}\label{eq47}
\eta(\xi) = \sum_{\mu=-\nu}^{\nu_1} p_{\mu} [g (\zeta)]^{\mu}
\end{equation}
is substituted in (\ref{eq46}). $p_\mu$ are coefficients and $g(\zeta)$ is solution of
simpler ordinary differential equation called simplest equation. Let the result
of this substitution be a polynomial of $g(\zeta)$. Eq. (\ref{eq47}) is a solution of Eq.(\ref{eq46}) 
if all coefficients of the obtained polynomial of $g(\zeta)$ are equal to $0$. This condition 
leads to a system of nonlinear algebraic equations. Each nontrivial solution of the last  system  leads to a solution of the studied  
nonlinear partial differential equation.
\par
 We search for solution of (45)  of kind $W=W(\zeta)={\sum\limits_{r=0}^q}  a_{r}g_{\zeta}^{r}$, where $\zeta = x -vt^{\prime}$, $a_{r}$ is a parameter, and $g(\zeta)$ is a solution of some ordinary differential equation, referred to as the simplest equation. The balance equation is $q=2m-2$. We assume that $m=3$, i.e. the  Abel equation of the first kind will play the role of simplest equation. Then
\begin{equation}\label{eq48}
W=a_{0}+a_{1}g+a_{2}g^2+a_{3}g_{3}+a_{4}g^4
\end{equation}
\begin{equation}\label{eq49}
\frac{dg}{d\zeta}=b_{0}+b_{1}g+b_{2}g^2+b_{3}g^3
\end{equation}
The solution of the differential equation of Abel can be given in a general form by the special
function introduced in \cite{v4} as follows. The differential equation of Abel can be written as
\begin{equation}\label{eq50}
\left(\frac{dg}{d \zeta} \right)^2 = c_0 +c_1g+c_2g^2+c_3g^3+c_4g^4+c_5g^5+c_6g^6
\end{equation}
where
\begin{eqnarray}\label{eq51}
c_0=b_0^2; \ c_1=2_0 b_1; \ c_2=2b_0 b_2 + b_1^2; \ c_3=2 b_0 b_3 + 2b_1 b_2; 
\nonumber \\ 
c_4=2b_1 b_3 + b_2^2; c_5=2 b_2 b_3 g^5; \ c_6=b_3^2
\end{eqnarray}
and its solution is given by the special function $V_{c_0,c_1,c_2,c_3,c_4,c_5,c_6}(\zeta;1,2,6)$.
In general the special function $V_{c_0,c_1;\dots,c_m}(\zeta;k,l,m)$ is the solution of the
ordinary differential equation
\begin{equation}\label{eq52}
\left(\frac{d^k g}{d \zeta^k} \right)^l = \sum \limits_{j=0}^m c_jg^j
\end{equation}
Then the solution of the Eq.(\ref{eq49}) is
\begin{equation}\label{eq53}
g(\zeta)=V_{b_0^2, 2_0 b_1, 2b_0 b_2 + b_1^2, 2 b_0 b_3 + 2b_1b_2, 2b_1 b_3 + b_2^2, 2 b_2 b_3 g^5, b_3^2}(\zeta;1,2,6)
\end{equation}
The relationships among the coefficients of the solution and the coefficients of the model are derived by solving a system of ten algebraic equations, and they are
\begin{eqnarray}\label{eq54}
a_{0}=-\frac{135 v b_3^2 + 64 b_2^2 \mu_2 (9 b_1 b_3 - 2 b_2^2) - 216b_2^2 \mu_2 b_1 b_3 + 540 \mu_2 b_1^2 b_3^2 + 28 \mu_2 b_2^4}{135 \mu_1 b_3^2},\nonumber \\ 
a_1 = - \frac{32 \mu_2 [-b_2^3 + 2b_2(9 b_1 b_3 - 2 b_2^2) + 27 b_1 b_2 b_3]}{45 \mu_1 b_3}, \ a_2 = - 16 \frac{\mu_2 (b_2^2 + 3 b_1 b_3)}{\mu_1}, \\
a_3=-64 \frac{\mu_2 b_2 b_3}{\mu_1}, \ a_4 = - 48 \frac{\mu_2 b_3^2}{\mu_1}, \
b_0 = \frac{b_2(9 b_1 b_3 - 2 b_2^2)}{27 b_3^2} \nonumber 
\end{eqnarray}
where $b_{1,2,3}$ are free parameters. Thus the solution of Eq.(\ref{eq45}) becomes
\begin{eqnarray}\label{eq55}
W(\zeta)=-\frac{135 v b_3^2 + 64 b_2^2 \mu_2 (9 b_1 b_3 - 2 b_2^2) - 216b_2^2 \mu_2 b_1 b_3 + 540 \mu_2 b_1^2 b_3^2 + 28 \mu_2 b_2^4}{135 \mu_1 b_3^2} - \nonumber \\
\frac{32 \mu_2 [-b_2^3 + 2b_2(9 b_1 b_3 - 2 b_2^2) + 27 b_1 b_2 b_3]}{45 \mu_1 b_3} V_{b_0^2, 2_0 b_1, 2b_0 b_2 + b_1^2, 2 b_0 b_3 + 2b_1b_2, 2b_1 b_3 + b_2^2, 2 b_2 b_3 g^5, b_3^2}(\zeta;1,2,6)- \nonumber \\
16 \frac{\mu_2 (b_2^2 + 3 b_1 b_3)}{\mu_1} V^2_{b_0^2, 2_0 b_1, 2b_0 b_2 + b_1^2, 2 b_0 b_3 + 2b_1b_2, 2b_1 b_3 + b_2^2, 2 b_2 b_3 g^5, b_3^2}(\zeta;1,2,6) -\nonumber \\
64 \frac{\mu_2 b_2 b_3}{\mu_1} V^3_{b_0^2, 2_0 b_1, 2b_0 b_2 + b_1^2, 2 b_0 b_3 + 2b_1b_2, 2b_1 b_3 + b_2^2, 2 b_2 b_3 g^5, b_3^2}(\zeta;1,2,6) - \nonumber \\
48 \frac{\mu_2 b_3^2}{\mu_1} V^4_{b_0^2, 2_0 b_1, 2b_0 b_2 + b_1^2, 2 b_0 b_3 + 2b_1b_2, 2b_1 b_3 + b_2^2, 2 b_2 b_3 g^5, b_3^2}(\zeta;1,2,6) \nonumber \\
\end{eqnarray}

This solution can be expressed by elementary functions as $b_0 = b_2(b_1 - 2b_2^2/(9b_3))/(3b_3)$. Then the solution of the
Abel equation is
\begin{eqnarray}\label{eq56}
g(\zeta)= V_{b_0^2, 2_0 b_1, 2b_0 b_2 + b_1^2, 2 b_0 b_3 + 2b_1b_2, 2b_1 b_3 + b_2^2, 2 b_2 b_3 g^5, b_3^2}(\zeta;1,2,6) = \nonumber \\
\frac{\exp \left[\left(b_1 - \frac{b_2^2}{3 b_3} \right) \zeta  \right]}{\sqrt{C-\frac{ b_3}{ \left(b_1 - \frac{b_2^2}{3 b_3} \right)}\exp\left[ 2 \left(b_1 - \frac{b_2}{3 b_3} \right) \zeta \right]}} - \frac{b_2}{3 b_3}
\end{eqnarray}
and the solution (\ref{eq55}) becomes
\begin{eqnarray}\label{eq57}
W(\zeta)=-\frac{135 v b_3^2 + 64 b_2^2 \mu_2 (9 b_1 b_3 - 2 b_2^2) - 216b_2^2 \mu_2 b_1 b_3 + 540 \mu_2 b_1^2 b_3^2 + 28 \mu_2 b_2^4}{135 \mu_1 b_3^2} - \nonumber \\
\frac{32 \mu_2 [-b_2^3 + 2b_2(9 b_1 b_3 - 2 b_2^2) + 27 b_1 b_2 b_3]}{45 \mu_1 b_3} 
g(\zeta)- 16 \frac{\mu_2 (b_2^2 + 3 b_1 b_3)}{\mu_1} g^2(\zeta) - \nonumber \\
64 \frac{\mu_2 b_2 b_3}{\mu_1} g^3(\zeta) - 48 \frac{\mu_2 b_3^2}{\mu_1} g^4(\zeta) \nonumber\\
\end{eqnarray}
where $g(\zeta)$ is given by Eq.(\ref{eq56}). Returning to the primary consideration, the solution of (\ref{eq40}) takes the form:
\begin{equation}\label{eq58}
U=W(\zeta)-\int{\mu(\tau)d\tau}
\end{equation}
where
\begin{equation}\nonumber
\zeta=\xi-v\tau-\int{\left[\int{\mu(\tau)d\tau}-\mu_{4}(\tau)\right]d\tau}
\end{equation}

We note, that Eq. (\ref{eq58}) represents the wave propagation in blood-filled arteries when $\mu_{3}\rightarrow{0}$, i.e. at vanishing viscosity. Such a behaviour is permissible for the blood near the walls of large arteries \cite{Demiray2002}. Then, the solution (\ref{eq58}) implies propagation of solitary waves, where the wave amplitude is a constant. When $\mu_{3}$ is no zero, i.e. dissipation in the system exists, the wave amplitude will be variable with respect to the temporal variable $\xi$ \cite{Demiray2002a}. However, this case will be a subject of our further consideration.

\end{document}